\documentclass{sf2a-conf2021}
\usepackage{graphicx}
\usepackage{hyperref}
\usepackage[]{natbib}
\usepackage{epstopdf}

\def\BibTeX{{\rm B\kern-.05em{\sc i\kern-.025em b}\kern-.08em
    T\kern-.1667em\lower.7ex\hbox{E}\kern-.125emX}}
\bibpunct{(}{)}{;}{a}{}{,}  

\begin{document}

\TitreGlobal{SF2A 2021}


\title{Provenance of astronomical data}

\runningtitle{Provenance of astronomical data}

\author{M. Servillat}\address{Laboratoire Univers et Théories, Observatoire de Paris, Université PSL, CNRS, Université de Paris, 92190 Meudon, France; \email{mathieu.servillat@obspm.fr}}





\setcounter{page}{237}


\maketitle


\begin{abstract}
In the context of Open Science, provenance has become a decisive piece of information to provide along with astronomical data. Provenance is explicitly cited in the FAIR principles, that aims to make research data Findable, Accessible, Interoperable and Reusable. The IVOA Provenance Data Model, published in 2020, puts in place the foundations for structuring and managing detailed provenance information, from the acquisition of raw data, to the dissemination of final products. The ambition is to provide for each astronomical dataset a sufficiently fine grained and detailed provenance information so that end-users understand the quality, reliability and trustworthiness of the data. This would ensure that the Reusable principle is respected.
\end{abstract}



\section{Introduction}

The idea behind Open Science\footnote{~See the dedicated UNESCO web page: \url{https://en.unesco.org/science-sustainable-future/open-science}} is to allow scientific information, data and outputs to be more widely accessible (Open Access) and more reliably harnessed (Open Data) with the active engagement of all the stakeholders (Open to Society).
Open Science is defined as "an inclusive construct that combines various movements and practices aiming to make multilingual scientific knowledge openly available, accessible and reusable for everyone" \citep{unesco_draft_rec}.
The aim is "to increase scientific collaborations and sharing of information for the benefits of science and society, and to open the processes of scientific knowledge creation, evaluation and communication to societal actors beyond the traditional scientific community" \citep{unesco_draft_rec}.

Open science is a policy priority for the European Commission\footnote{~See the dedicated EU web page: \url{https://ec.europa.eu/info/research-and-innovation/strategy/strategy-2020-2024/our-digital-future/open-science_en}} and the standard method of working under its research and innovation funding programmes as it improves the quality, efficiency and responsiveness of research. One of the ambition of this policy is to build a European Open Science Cloud (EOSC), i.e. an environment that cuts across borders and scientific disciplines to store, share, process and reuse research digital objects (like publications, data, and software) that are Findable, Accessible, Interoperable and Reusable (FAIR).

In the astronomy domain, the FAIR principles \citep{Wilkinson2016} have been a matter of concern since more than 20 years, primarily within the International Virtual Observatory Alliance\footnote{~\url{https://www.ivoa.net}} (IVOA), that provides standards to foster interoperability and enable the production of Open Data. Several astronomical research infrasctructures are involved in the European Horizon 2020 ESCAPE project\footnote{~\url{https://projectescape.eu}} (European Science Cluster of Astronomy \& Particle physics ESFRI research infrastructures) that brings together the astronomy, astroparticle and particle physics communities and puts together a cluster with aligned challenges of data-driven research.

The Virtual Observatory ecosystem already provides robust solutions to Find and Access astronomical data in an Interoperable way. However, the Reusable principle is more subjective and requires dedicated rich metadata to demonstrate the quality, reliability and trustworthiness of the data. Detailed and structured provenance information is then key information to provide along with the astronomical data.

In this context, and with the implication of several members of the French ASOV (\textit{Action Sp\'ecifique Observatoire Virtuel}), a series of meetings and workshops took place over the last years to model provenance information \citep{2020ivoa.spec.0411S} and implement related tools \citep{B9-56_adassxxx}. A provenance management system has then been proposed for astronomical facilities \citep{10.1007/978-3-030-80960-7_20}.

\section{Requirement for structured provenance}
\label{req_struct_prov}

There are clear advantages to retain provenance information as structured, machine-readable data, in particular in the context of Open Science:

\vspace{-.5\topsep}
\begin{itemize}
    \setlength{\parskip}{2pt}
    \item \textbf{Quality / Reliability / Trustworthiness} of the products: the simple fact of being able to show its provenance is sufficient to give more value to a product, and if the provenance information is detailed, the value is higher.
    \item \textbf{Reproducibility requirement} in many projects: provenance details are essential to be able to rerun each activity (maybe testing and improving each step); Having this information, it may not be necessary to keep every intermediate file that is easily reproducible (hence a possible gain on storage space and costs).
    \item \textbf{Debugging}: with detailed provenance, it is not necessary to restart from scratch, as one can locate in the provenance graph the faulty parts or the products to be discarded, and reprocess only from the identified failing steps.
\end{itemize}
\vspace{-.5\topsep}

We often realize too late that there are missing elements or links in the provenance. The capture of the provenance should thus be as detailed as possible. It should also be as naive as possible: provenance should trace what happened, which is different to the workflow approach where one anticipates what should happen. The good practice would thus be to record provenance events directly when they occur, with the relevant links to what happened before, and not considering what will happen after.

\section{A Provenance data model}

The IVOA validated in April 2020 a Provenance Data Model \citep{2020ivoa.spec.0411S} to structure the provenance information. It is based on the World Wide Web Consortium (W3C) PROV core concepts of Entity, Activity and Agent \citep{std:W3CProvDM} with a dedicated set of classes for the activity description (e.g. method, algorithm, software) and the activity configuration (e.g. parameters).

Provenance is related by definition to the origin of a product (where does it come from?), but also the path followed to generate this product (what has been done?).
Provenance is thus seen as a chain of activities and entities, used and generated. With the core data model, basic objectives are achieved: use of unique identifiers, traceability of the operations, connection with contacts for further information, citation or acknowledgement. By following the full IVOA data model, more advanced questions are answered: What happened during each activity? How was the activity tuned to be executed properly? What kind of content is in the entities?

The data model is a basis for the development of tools and services, see e.g.: \citet{P9-89_adassxxx, P9-250_adassxxx, P9-216_adassxxx, 2020ASPC..522..199S, 2020ASPC..522..545S}.

\begin{figure}[ht!]
  \centering
  \includegraphics[width=\textwidth,clip]{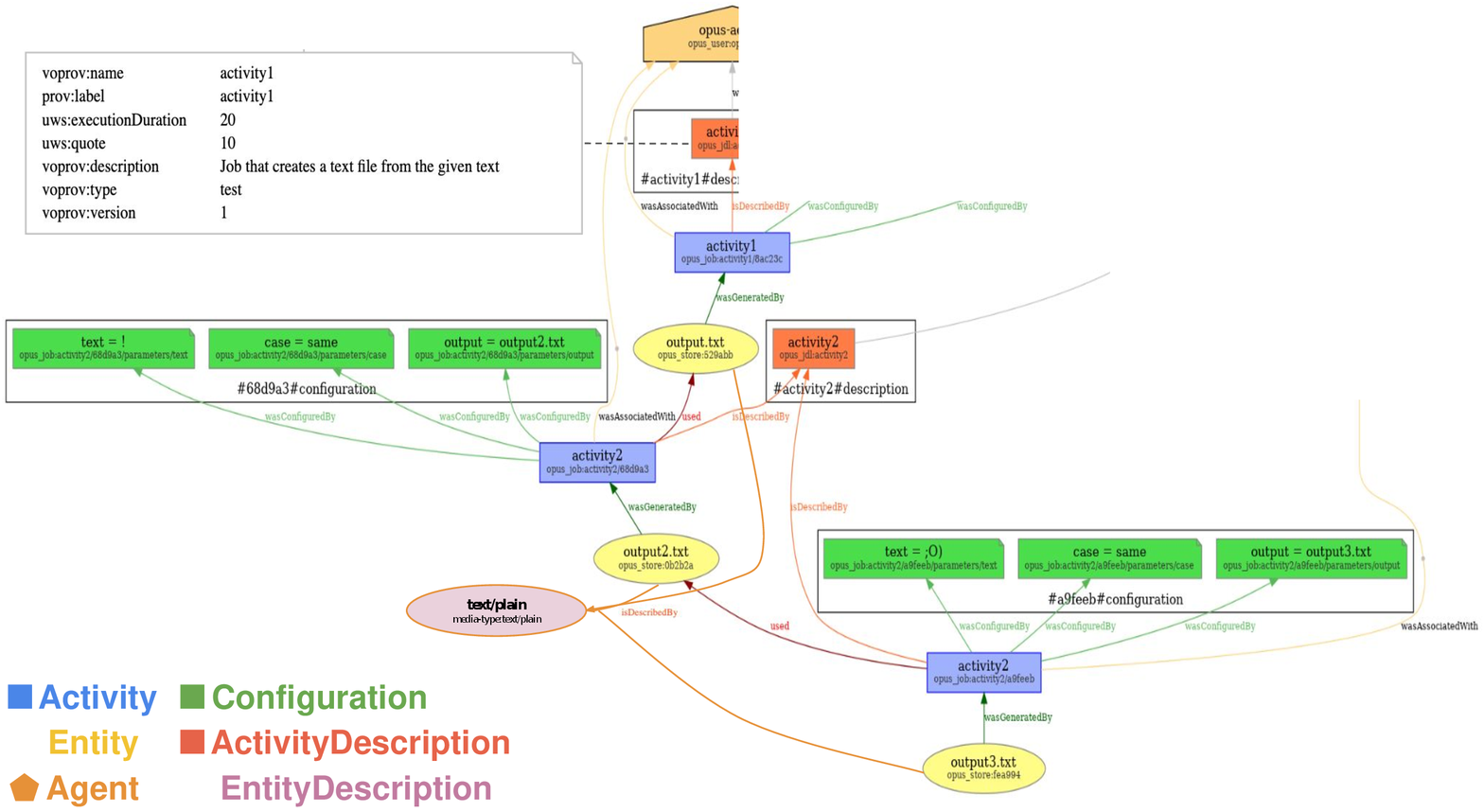}
  \caption{Example of an IVOA Provenance graph using the main concepts of the IVOA Provenance Data Model \citep{2020ivoa.spec.0411S}. The graph shows a sequence of 3 activities. Each activity performs an operation for which the description and the configuration are available and explicit.}
  \label{prov:fig1}
\end{figure}

The data model and related implementations provide a standard formalism to write and exchange the provenance information. This is illustrated in Figure~\ref{prov:fig1}, where the \texttt{voprov} Python package\footnote{~\url{https://github.com/sanguillon/voprov}} was used to generate a graph of three activities, executed with the OPUS job manager \citep{P9-89_adassxxx}.

\section{Provenance in practice}

\subsection{Full provenance}

It is tempting to limit the provenance information to a list of keywords associated to a data product. However, the full provenance is to be seen as a global graph of activities and entities up to the raw data, which cannot be embedded in the entities themselves. This consideration led to the development of an advanced provenance management system, with the concepts of capture "inside" (i.e. during the execution of a processing pipeline), storage of all provenance events in a central database, and visualization and exploration of the full provenance through database queries  \citep[for more details, see ][]{10.1007/978-3-030-80960-7_20}.

To ease the capture of provenance inside a pipeline, a Python package, \texttt{logprov}\footnote{~\url{https://github.com/mservillat/logprov}}, is in development along with the pipelines and science tools developed for the Cherenkov Telescope Array (CTA). This capture tool was initially implemented for \texttt{gammapy}\footnote{~\url{https://github.com/gammapy/gammapy}} and its high level interface. The usage of \texttt{logprov} requires to insert decorators before the Python functions or classes one wish to trace. Provenance events are then written to a structured log. In addition, a definition file of the activities can be added to record more detailed descriptions of the activities and entities. To make this capture efficient, it is highly recommended to structure the code and pipelines in well defined functions, with identified inputs and outputs (i.e. not contained in local variables, but globally accessible).

\subsection{Last-step embedded provenance}

Along with the definition of the full provenance, the idea of an optimized subset of provenance that could be embedded in an entity has emerged. We defined the last-step provenance as a minimal list of keywords that gives information on the last activity (general process/workflow, software versions, contacts...), including links to used and generated entities \citep{B9-56_adassxxx}. Such a list is a restriction of the full provenance information, that can be stored in a file header (e.g. using the FITS file format) or a flat table.

\begin{figure}[ht!]
  \centering
  \vspace{-0.6cm}
  \includegraphics[width=.6\textwidth,clip]{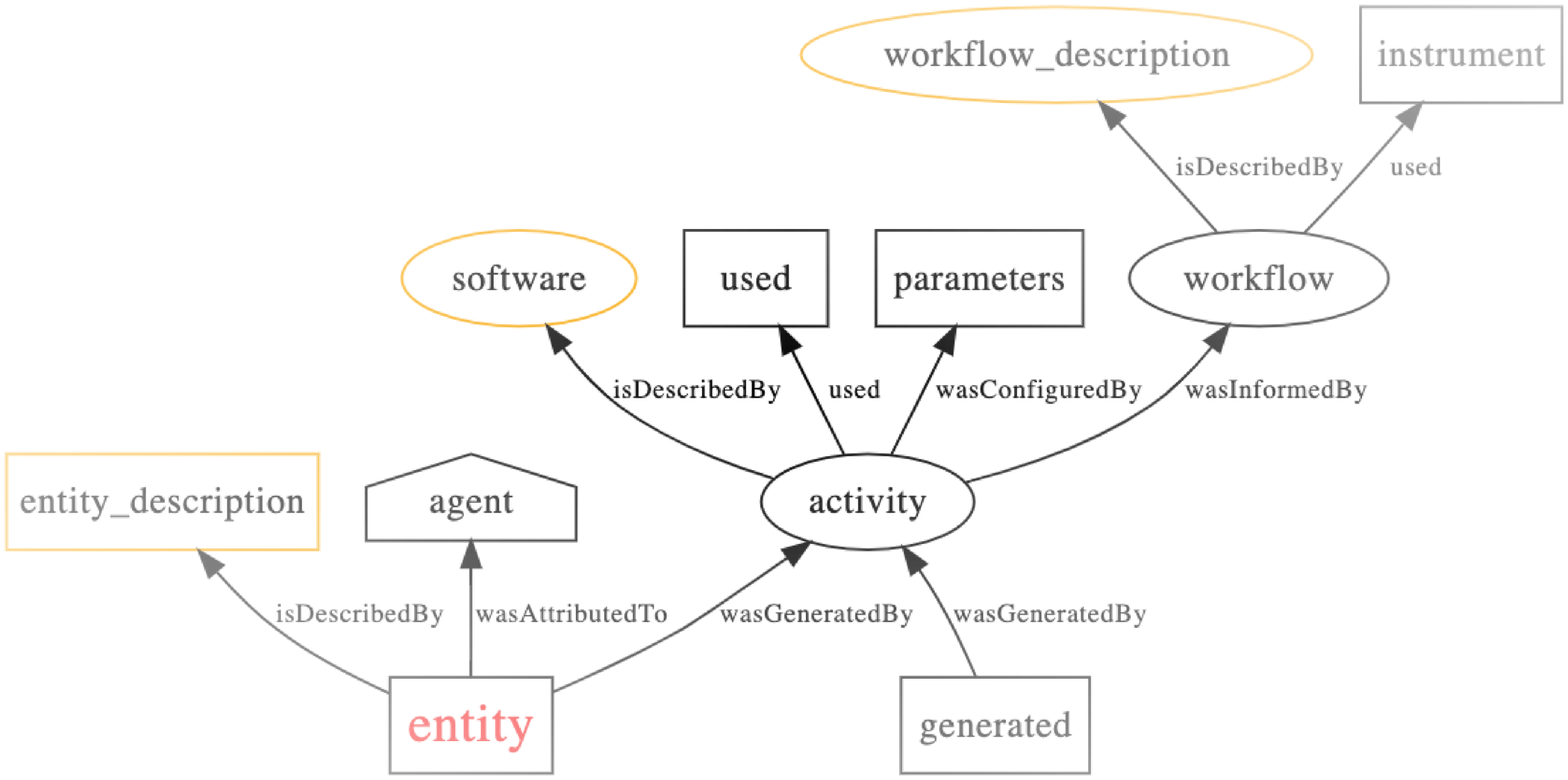}
  \vspace{-0.6cm}
  \caption{Provenance graph template that defines the prefixes of last-step provenance attributes.}
  \label{prov:fig2}
\end{figure}

The last-step provenance is composed of attributes that refer to several items in a provenance graph basic template. For example, the entity itself is described by attributes like \texttt{entity\_id}, \texttt{entity\_location}...
The template graph is shown in Figure~\ref{prov:fig2}. The main node in the graph is the entity that transports this last-step provenance, which is attached to the activity that generated the entity. The context is provided by attributes that describe the entity (e.g. \texttt{entity\_content\_type}) and the activity (e.g. \texttt{software\_name}, \texttt{software\_version, \texttt{software\_docurl}}), and parameters that configure the activity. The activity may be part of a general workflow, itself described, and maybe linked to an instrument. Finally, the identifiers of used entities, and the generated ones during the same process allow for the exploration of the previous, or next, or parallel steps. By resolving those identifiers and combining chained last-step provenance records, one could thus explore or reconstruct the full provenance.

\subsection{Provenance access protocols}

Two access protocols are being discussed within IVOA, in order to be able to query the provenance information:

\vspace{-.5\topsep}
\begin{itemize}
    \setlength{\parskip}{2pt}
    \item \textbf{ProvSAP}: a Simple Access Protocol that returns a W3C PROV file from a regular GET query on an HTTP endpoint, where the main argument is ID with the identifier of the entity or activity for which the provenance graph is queried. This system is for example implemented in the OPUS job manager \citep{P9-89_adassxxx} and in other tools \citep{2020ASPC..522..545S}.
    \item \textbf{ProvTAP}: IVOA Table Access Protocol (TAP) using a schema based on the IVOA Provenance data model \citep{2019ASPC..523..313B}. It's a reverse mechanism to locate data through queries on its provenance. This approach also enables queries to test the data quality, based on the analysis of parameters of activities.
\end{itemize}

The solutions developed here for provenance management (full provenance capture and storage, last-step embedded provenance, and provenance access protocols) thus provide several efficient and standardized approaches that can be adapted to various astronomy projects of different sizes.

\begin{acknowledgements}
We acknowledge support from the ESCAPE project funded by the EU Horizon 2020 research and innovation program (Grant Agreement n.824064). Additional funding was provided by the INSU (Action Sp\'ecifique Observatoire Virtuel, ASOV), the Action F\'ed\'eratrice CTA at the Observatoire de Paris and the Paris Astronomical Data Centre (PADC).
\end{acknowledgements}

\bibliographystyle{aa}  
\bibliography{servillat_S0} 

\begin{thebibliography}{12}
\expandafter\ifx\csname natexlab\endcsname\relax\def\natexlab#1{#1}\fi

\bibitem[{{Bonnarel} {et~al.}(2019){Bonnarel}, {Louys}, {Mantelet},
  {Nullmeier}, {Servillat}, {Riebe}, \& {Sanguillon}}]{2019ASPC..523..313B}
{Bonnarel}, F., {Louys}, M., {Mantelet}, G., {et~al.} 2019, in ASP Conf. Ser.,
  Vol. 523, ADASS XXVII, ed. P.~J. {Teuben}, M.~W. {Pound}, B.~A. {Thomas}, \&
  E.~M. {Warner}, 313

\bibitem[{Landais {et~al.}(2021)Landais, Servillat, Bonnarel, Louys,
  Sanguillon, \& Michel}]{P9-216_adassxxx}
Landais, G., Servillat, M., Bonnarel, F., {et~al.} 2021, in ASP Conf. Ser.,
  Vol. TBD, ADASS XXX, ed. J.-E. {Ruiz} \& F.~{Pierfederici}, TBD

\bibitem[{Moreau {et~al.}(2013)Moreau, Missier, Belhajjame, B'Far, Cheney,
  Coppens, Cresswell, Gil, Groth, Klyne, Lebo, McCusker, Miles, Myers, Sahoo,
  \& Tilmes}]{std:W3CProvDM}
Moreau, L., Missier, P., Belhajjame, K., {et~al.} 2013, {PROV-DM}: The PROV
  Data Model, {W3C Recommendation}

\bibitem[{Sanguillon {et~al.}(2021)Sanguillon, Arrabito, Boisson, Bregeon,
  Kosack, \& Servillat}]{P9-250_adassxxx}
Sanguillon, M., Arrabito, L., Boisson, C., {et~al.} 2021, in ASP Conf. Ser.,
  Vol. TBD, ADASS XXX, ed. J.-E. {Ruiz} \& F.~{Pierfederici}, TBD

\bibitem[{{Sanguillon} {et~al.}(2020){Sanguillon}, {Bonnarel}, {Louys},
  {Nullmeier}, {Riebe}, \& {Servillat}}]{2020ASPC..522..545S}
{Sanguillon}, M., {Bonnarel}, F., {Louys}, M., {et~al.} 2020, in ASP Conf.
  Ser., Vol. 522, ADASS XXVII, ed. P.~{Ballester}, J.~{Ibsen}, M.~{Solar}, \&
  K.~{Shortridge}, 545

\bibitem[{Servillat {et~al.}(2021{\natexlab{a}})Servillat, Aicardi, Cecconi, \&
  Mancini}]{P9-89_adassxxx}
Servillat, M., Aicardi, S., Cecconi, B., \& Mancini, M. 2021{\natexlab{a}}, in
  ASP Conf. Ser., Vol. TBD, ADASS XXX, ed. J.-E. {Ruiz} \& F.~{Pierfederici},
  TBD

\bibitem[{{Servillat} {et~al.}(2020{\natexlab{a}}){Servillat}, {Boisson},
  {Lefaucheur}, {Kosack}, {Sanguillon}, {Louys}, \&
  {Bonnarel}}]{2020ASPC..522..199S}
{Servillat}, M., {Boisson}, C., {Lefaucheur}, J., {et~al.} 2020{\natexlab{a}},
  in ASP Conf. Ser., Vol. 522, ADASS XXVII, ed. P.~{Ballester}, J.~{Ibsen},
  M.~{Solar}, \& K.~{Shortridge}, 199

\bibitem[{Servillat {et~al.}(2021{\natexlab{b}})Servillat, Bonnarel, Boisson,
  Louys, Ruiz, \& Sanguillon}]{10.1007/978-3-030-80960-7_20}
Servillat, M., Bonnarel, F., Boisson, C., {et~al.} 2021{\natexlab{b}}, in
  Provenance and Annotation of Data and Processes, ed. B.~Glavic,
  V.~Braganholo, \& D.~Koop (Cham: Springer International Publishing), 244--249

\bibitem[{Servillat {et~al.}(2021{\natexlab{c}})Servillat, Bonnarel, Louys, ,
  \& Sanguillon}]{B9-56_adassxxx}
Servillat, M., Bonnarel, F., Louys, M., , \& Sanguillon, M. 2021{\natexlab{c}},
  in ASP Conf. Ser., Vol. TBD, ADASS XXX, ed. J.-E. {Ruiz} \&
  F.~{Pierfederici}, TBD

\bibitem[{{Servillat} {et~al.}(2020{\natexlab{b}}){Servillat}, {Riebe},
  {Boisson}, {Bonnarel}, {Galkin}, {Louys}, {Nullmeier}, {Renault-Tinacci},
  {Sanguillon}, \& {Streicher}}]{2020ivoa.spec.0411S}
{Servillat}, M., {Riebe}, K., {Boisson}, C., {et~al.} 2020{\natexlab{b}}, {IVOA
  Provenance Data Model Version 1.0}, IVOA Recommendation

\bibitem[{UNESCO(2021)}]{unesco_draft_rec}
UNESCO, Director-General, A.~A. 2021, UNESCO Circular Letter: Draft text of the
  UNESCO Recommendation on Open Science,
  \url{https://unesdoc.unesco.org/ark:/48223/pf0000378381}

\bibitem[{Wilkinson {et~al.}(2016)Wilkinson, Dumontier, Aalbersberg, Appleton,
  Axton, Baak, Blomberg, Boiten, da~Silva~Santos, Bourne,
  {et~al.}}]{Wilkinson2016}
Wilkinson, M.~D., Dumontier, M., Aalbersberg, I.~J., {et~al.} 2016, Scientific
  Data, 3

\end{thebibliography}

\end{document}